\documentclass[twocolumn,prl,aps,floatfix,showpacs]{revtex4}%
\usepackage{graphicx}
\usepackage{bm}
\usepackage{amsmath}
\usepackage{amsfonts}
\usepackage{amssymb}%
\providecommand{\U}[1]{\protect\rule{.1in}{.1in}}
\begin{document}
\title{Effect of dimensionality on the charge-density-wave in few-layers 2H-NbSe2}
\author{Matteo Calandra$^{1}$}
\author{I. I. Mazin$^{2}$}
\author{Francesco Mauri$^{1}$}
\affiliation{$^{1}$CNRS and Institut de Min\'eralogie et de Physique des Milieux
condens\'es, case 115, 4 place Jussieu, 75252, Paris cedex 05, France}
\affiliation{$^{2}$Naval Research Laboratory, 4555 Overlook Ave. SW, Washington, DC 20375}
\date{\today}

\begin{abstract}
We investigate the charge density wave (CDW) instability in single and double
layers, as well as in the bulk 2H-NbSe$_{2}$. We demonstrate that the density
functional theory correctly describes the metallic CDW state in the bulk
2H-NbSe$_{2}$. We predict that both mono- and bilayer NbSe$_{2}$ undergo a CDW
instability. However, while in the bulk the instability occurs at a momentum
$\mathbf{q}_{CDW}\approx\frac{2}{3}\mathbf{\Gamma M}$, in free-standing
layers it occurs at $\mathbf{q}_{CDW}\approx\frac{1}{2}\mathbf{\Gamma M}$.
Furthermore, while in the bulk the CDW leads to a metallic state, in a
monolayer the ground state becomes semimetallic, in agreement with recent
experimental data. We elucidate the key role that an enhancement of the
electron-phonon matrix element at $\mathbf{q}\approx\mathbf{q}_{CDW}$ plays in
forming the CDW ground state.

\end{abstract}

\maketitle

Charge density wave (CDW) is one of the most common and most intriguing
phenomena in solid state physics\cite{book}. The concept originated in the
seminal paper by Peierls\cite{Peierls} pointing out a divergence in the 1D
response functions at a wave vector equal to twice the Fermi vector. Later
this concept was generalized onto the 2D and 3D system with \textquotedblleft
nesting\textquotedblright, that is, quasi-1D portions of the Fermi surface. 
On the other hand, the fact that the Peierls instability is logarithmic,
and therefore very fragile, has led to the question of whether the actual CDW
observed in quasi-2D materials are indeed manifestations of the Peierls
instability.\cite{Johannes08}

In the last decades the role of nesting in various structural and magnetic
instabilities has been wildly debated, particularly with respect to the most
venerable CDW materials, NbSe$_{2}$ and related dichalcogenides. It has been
pointed out that the bare susceptibility does not have any sharp peak at the
CDW wave vector, $\mathbf{q}_{CDW}=(2\pi/3,2\pi/3),$ but at best a broad and
shallow peak\cite{Doran78a,Johannes06,Johannes08} in the real part, while the
imaginary part (which directly reflects the Fermi surface nesting) does not peak at
$\mathbf{q}_{CDW}$ at all\cite{Whango92,Johannes06,Johannes08}.

This suggests that momentum dependence of the electron-phonon interaction
plays a crucial role in driving CDW
instabilities\cite{Doran78b,Johannes06,Johannes08}. An indirect confirmation
comes from neutron \cite{Moncton75_77} and X-ray scattering \cite{Murphy05_08}
phonon-dispersion data where a marked softening was detected in only one of
the low energy modes close to $\mathbf{q}_{CDW}$.

A very clean test for the described hypothesis are purely 2D systems ---
mono- and bilayers of the same compounds. Lacking $k_{z}$ dispersion, they
should be much more sensitive to the Fermi surface nesting than their 
3D counterparts.
2H-NbSe$_{2}$ is particularly promising as its 2D structures have been
synthesized, and found to show a curious semimetallic behavior
\cite{Novoselov05}. Interestingly the conductivity of a NbSe$_2$ layer 
in a field-effect transistor as a function of the gate
voltage ($V_{g}$), which controls the doping, 
is very different from that of good metallic monolayers.
While typical metals have a conductivity rather weakly dependent on doping,
NbSe$_{2}$ monolayers show a strong monotonic dependence on $V_{g}$. Such a
large variation (more then a factor of two in the -70 to +70 V $V_g$-range) is
characteristic of semiconductors, semimetals as graphene, 
or metals with a pseudogap at the Fermi level.
The first principles calculations predict a good metal for a
monolayer in the undistorted bulk structure \cite{Lebegue09}, in disagreement
with experiments, which suggest a possibility of a semimetallic band
structure.

In this Letter we study the CDW instability in the bulk 2H-NbSe$_2$, in the 
bilayer and in the monolayer NbSe$_2$, both by calculating the phonon
spectra in the high-symmetry phase and by full structural optimization\cite{footnote}. 
We find that all three studied structures are unstable against a CDW formation
with, however, \textit{different} patterns, despite the extreme similarity of
the Fermi surfaces. We then demonstrate that the crucial difference comes from the
\textbf{q-}dependent electron-phonon matrix elements, rather than directly from
the Fermi surface (FS). 
Finally we show that the CDW distorted monolayer is semimetallic and
the large variation of its conductivity as a function of $V_{g}$ detected in
experiments is a manifestation of the CDW.

We simulate a single layer or a bilayer, as usually, by introducing a thick
vacuum layer (of the order of 10 \AA ). We then performed the geometrical
optimization in one unit cell, without a CDW. We found that the distance
between the Nb and Se planes barely changes: from 1.67 \AA~ in the bulk to
1.69 \AA~ in the bilayer (same in the monolayer). On the contrary, the Nb-Nb
interplanar distance in the bilayer is significantly enlarged compared to the
bulk (6.27 \AA \ $vs.$ 6.78 \AA ).
 \begin{figure}[t]
\includegraphics[width=.9\columnwidth]{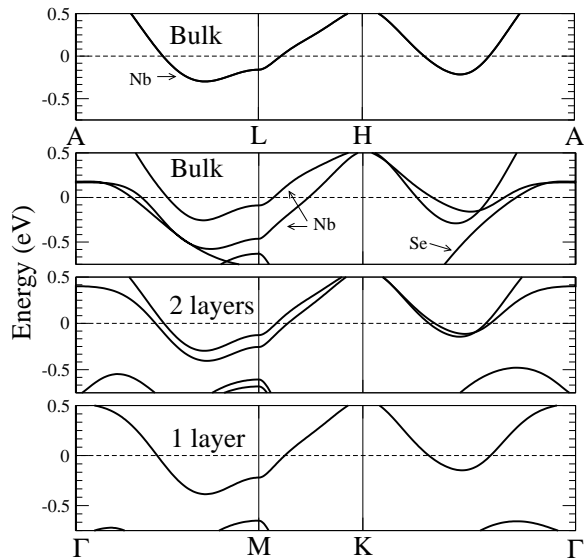}\caption{Electronic
structure of the undistorted bulk, bilayer and monolayer 2H-NbSe$_{2}$. }%
\label{fig:bands}%
\end{figure}

The pseudopotential electronic structure of the NbSe$_2$ bilayers and monolayers
is compared with that of the bulk NbSe$_{2}$ in Fig. \ref{fig:bands} (LAPW
calculations yield very similar results). For the bulk case, as known from
previous calculations \cite{Mattheiss73,Johannes06}, three bands cross the
Fermi level $(\epsilon_{f})$. One is the antibonding Se $p_{z}$ bands, which
is highly dispersive along the z-axis and does not cross $\epsilon_{f}$ in the
A-L-H plane. The other two are formed by Nb d-states. These two bands are
degenerate in the A-L-H plane in the scalar relativistic calculations
(spin-orbit interaction removes the degeneracy) while the degeneracy is 
lifted for other values of $k_{z}$ \cite{Johannes06} by the interlayer
interaction. As a result, while one of the FSs originated from Nb d-bands is
two dimensional, the other acquires considerable $k_{z}$ dispersion, so that a
model neglecting hopping between the layers does not apply \cite{Inosov08}.

\begin{figure}[t]
\includegraphics[width=0.41\columnwidth]{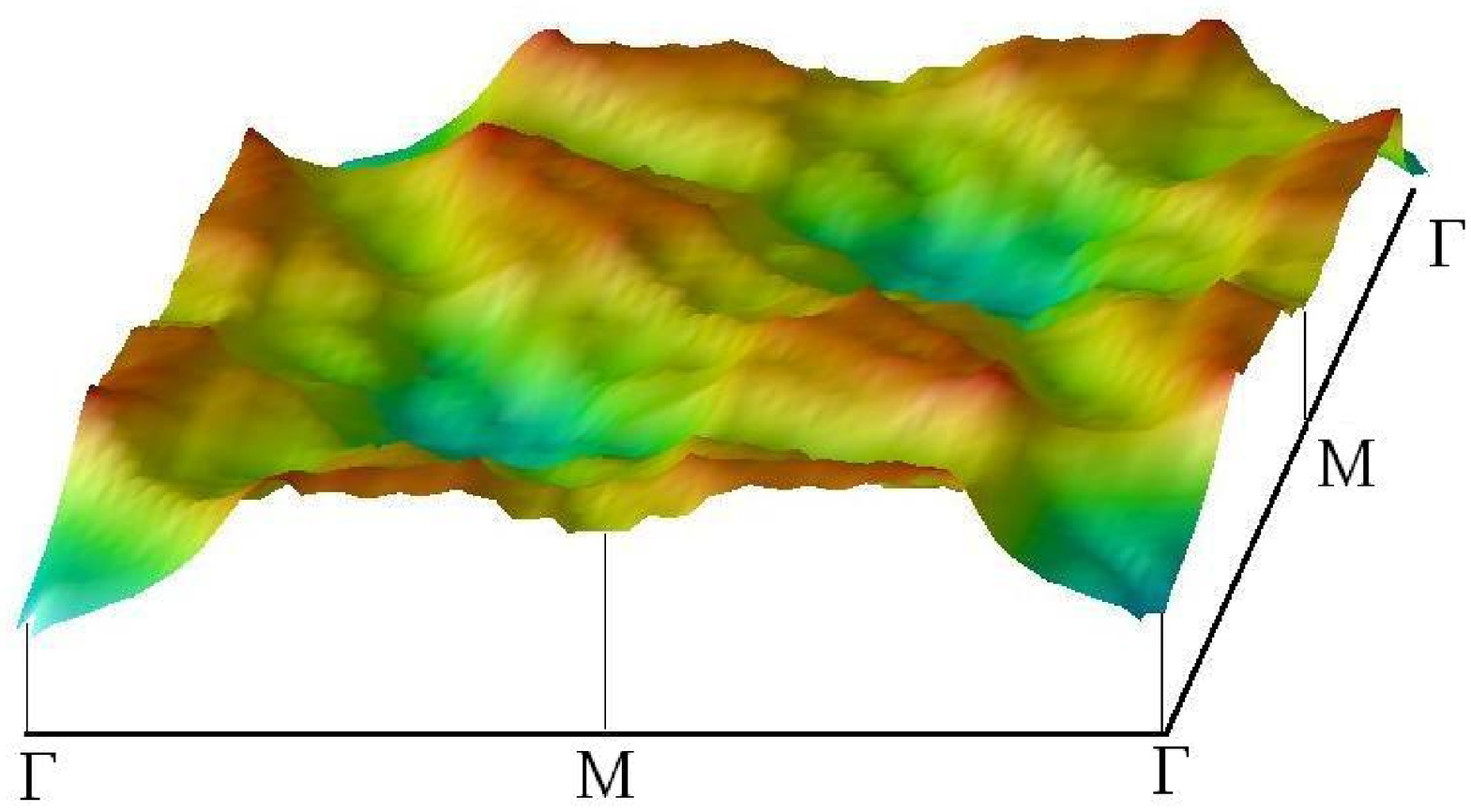}
\includegraphics[width=0.45\columnwidth]{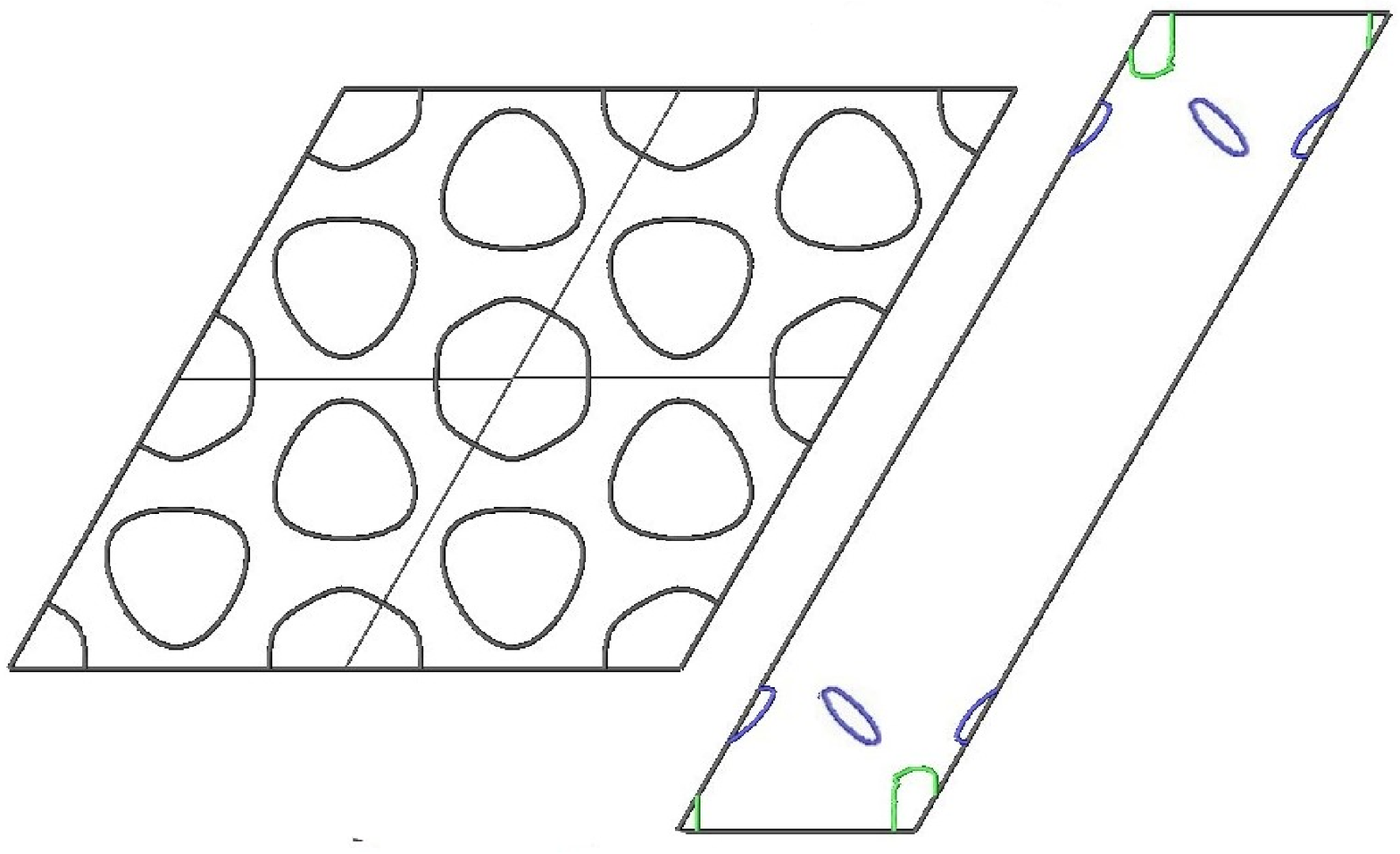}
\caption{(color online) Left: Real
part of the bare static electronic susceptibility (in arbitrary
units) with constant matrix elements, same as the real part of the phonon
self-energy $\Pi_{\sigma,\mu}^{^{\prime}}(\mathbf{q},\omega=0)$, Eq.
(\protect\ref{Pi}), calculated using the LAPW band structure of a
single layer NbSe$_{2}$. Center: The corresponding FS. Four Brillouin zones are shown, $\Gamma$ points are
at the grid nodes. Right: The same, for the resulting 4$\times$1 CDW. }%
\label{fig:monofs}%
\end{figure}In a bilayer, the Se-Se interplanar interaction is reduced, while
in the monolayer it is absent. As a result, the Se band sinks below the Fermi
level, and only Nb the derived Fermi surfaces are present, and they are strictly
2D, that is, seemingly more prone to the nesting effects. The calculated FS
for a single layer (Fig. \ref{fig:monofs}) consists of a rounded hexagon at
$\Gamma$ and two rounded triangles at $\mathbf{K}$, just as the Nb FS pockets
in the bulk 2H-NbSe$_{2}$\cite{Johannes06}.
  
\begin{figure}[ptb]
\includegraphics[width=.9\columnwidth]{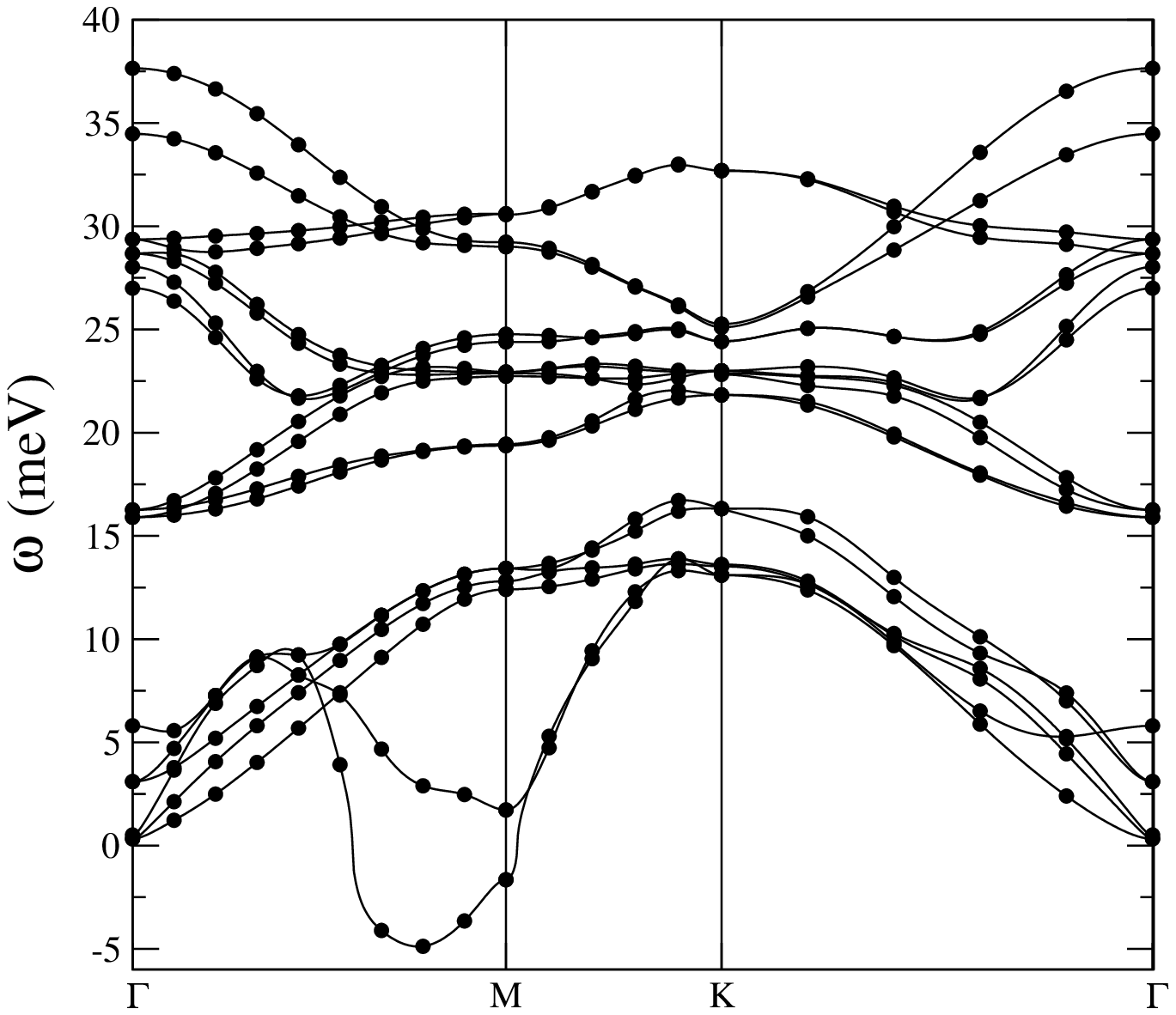}
\includegraphics[width=.9\columnwidth]{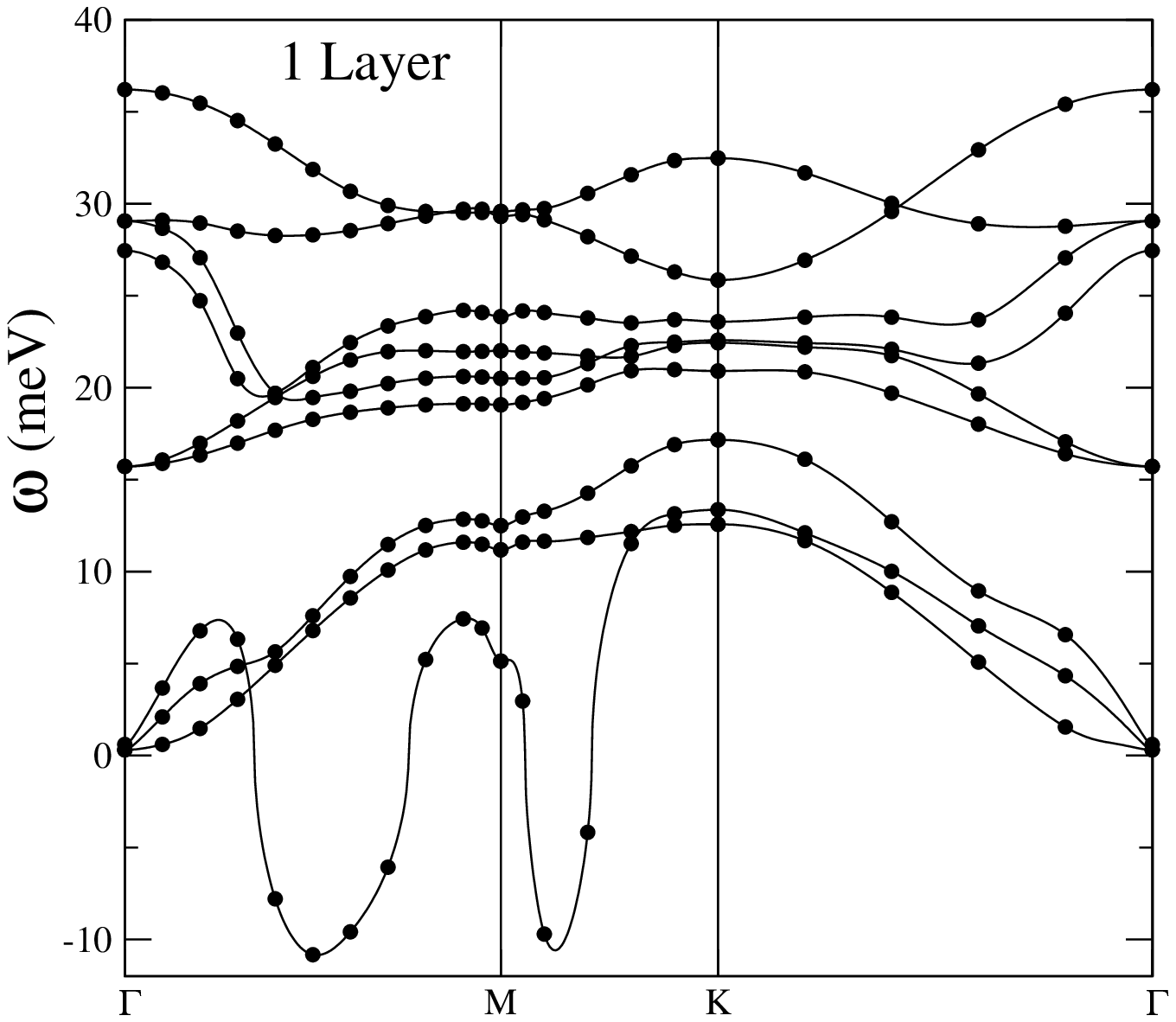}\caption{(color
online) Phonon dispersion of bulk (top) and monolayer (bottom) 2H-NbSe$_{2}$.
Each point corresponds to a linear response calculations. The spline connecting
the points is guide to the eye.}%
\label{fig:phonon}%
\end{figure}

Next, we have computed the phonon dispersions for the bulk and single-layer
NbSe$_{2}$. In Fig. \ref{fig:phonon} we plot the  phonon dispersions in the
monolayer (bulk) calculated using a $40\times40$ ($20\times20\times6$ )
k-points grid and Fermi temperature $\tau=68$ meV ($\tau=270$ meV) with
Hermite-Gaussian distributions 
(See Ref. \cite{CalandraCaC6} for more details.). In the bulk,
the highest energy acoustic mode is unstable at approximately $\mathbf{q}%
_{CDW}=2/3\Gamma$M, in agreement with the experiment.%\cite{Wilson74}.
 These
calculations implicitly include the renormalization of the bare phonon
frequencies due to the electron-phonon interaction, as described by the real
part of the phonon self-energy of a phonon mode $\mu$, namely
\begin{equation}
\Pi_{\tau,\mu}^{\prime}(\mathbf{q},\omega=0)\sim\sum_{\mathbf{k},nm}%
\frac{|g_{\mathbf{k}i,\mathbf{k+q},j}^{\tau,\mu}|^{2}(f_{\mathbf{k+q},j}%
^{\tau}-f_{\mathbf{k}i}^{\tau})}{\epsilon_{\mathbf{k+q},j}^{\tau}%
-\epsilon_{\mathbf{k},i}^{\tau}},\label{Pi}%
\end{equation}
where $\epsilon_{\mathbf{k}i}$ is the one-electron energy, $g_{\mathbf{k}%
i,\mathbf{k+q},j}^{\mu}$ is the electron-phonon matrix element and
$f_{\mathbf{k}i}^{\tau}$ is the Fermi function. 
If one neglects the \textbf{k} dependence of the matrix elements in Eq. \ref{Pi}, this
expression becomes proportional to the real part of the unrenormalized static
electronic susceptibility, $\chi_{0}^{\prime}(\mathbf{q)}$; as shown in Ref.
\cite{Johannes06}, in NbSe$_{2}$ $\chi_{0}^{\prime}(\mathbf{q)}$ has a broad
maximum at $\mathbf{q}_{CDW}=2/3\Gamma$M, even though the imaginary part
(which directly reflects the FS nesting) does not.

Thus, DFT correctly describes the CDW instability in the bulk NbSe$_{2}.$ With
this in mind, we move to a NbSe$_{2}$ monolayer and observe that (a) now the
most unstable phonon mode along $\Gamma$M appears near (1/2)$\Gamma$M
$=$(0.5$\pi/a,0)$ and (b) the instability expands over a substantially larger
k-point region, as phonon are unstable also along MK. We also looked at the
bare susceptibility $\chi_{0}^{\prime}$ and found a strip-like region of
enhanced values, extending from 0.4$\Gamma$M to 0.75$\Gamma$M, with a width of
approximately 0.15 of the $\Gamma$M vector (0.15$\pi/a).$ This is in agreement
with the instabilities found in the linear-response phonon calculations along
$\Gamma$M and MK.

\begin{figure}[ptb]
\includegraphics[width=0.9\columnwidth]{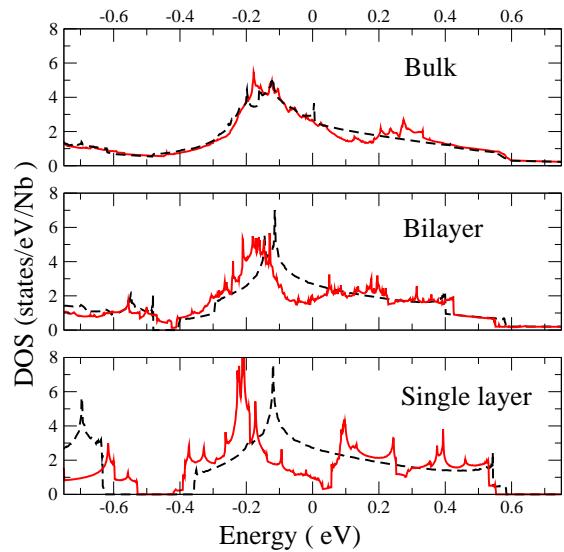}\caption{(color online):
Pseudopotentials electronic density of states for undistorted (black-dashed)
and distorted (red) bulk, bilayer and monolayer 2H-NbSe$_{2}$.}%
\label{fig:dos}%
\end{figure}

To understand which electronic states are involved in the CDW formation we
have repeated the calculations with an increased electronic temperature $\tau$.
%(larger temperature probes the electronic states farther removed from the Fermi surface).
We found that the phonon frequency of the highest acoustic
mode strongly depends on $\tau$ for $\tau > 0.3 $eV.  The instability occurs
for $\tau \approx 0.3$ eV, and for smaller temperatures the (imaginary) 
phonon frequency of the unstable mode is essentially constant. 
This means that electronic states within a 0.3 eV window around
$\epsilon_{f}$ are responsible for the CDW instability, which is
not a Fermi surface effect\cite{RiceScott75,Wilson77, Doran78a,Straub99}, 
but involves states in the full Nb-derived d-band
(which is only $1$ eV wide, see Fig. \ref{fig:dos}). 
As discussed above, $\chi_{0}^{\prime}$ does have a maximum at
\textbf{q}$_{CDW},$ but this maximum is relatively weak and broad, and, most
importantly, weakly dependent on $\tau$. Thus, the CDW  
is mainly the result of an enhancement of
the electron-phonon matrix element close to $\mathbf{q}_{CDW}$. 
The self-consistent screening of the electron-phonon matrix element as a function of
$\tau$ is crucial to describe the CDW. 

The next test is to take a superstructure suggested by the calculated CDW
vector [($\pi/3,0)$ or ($\pi/4,0)]$ and to optimize the cell dimensions and the
atomic positions. 
%Besides these two superstructures, $3\times1$ and
%$4\times1,$ for a monolayer we have also checked others: $2\times1$,
%$2\times2$, and $3\times3$.
 Each time we started from slightly randomized
atomic positions and minimized the total energy. We found that in the bulk the
$4\times1$ and $3\times1$ supercells both converge to a lower energy than the
undistorted cell, consistent with the calculated phonon spectra in Fig.
\ref{fig:phonon}, the $3\times1$ being lower in energy (with the energy gain
of 0.1 mRyd/Nb with respect to the undistorted structure). The other
supercells converged to the undistorted structure. The distortion of the
$3\times1$ supercell is consistent with the $\Sigma_{1}$ phonon pattern
detected in inelastic neutron scattering \cite{Moncton75_77}, involving not
only an in-plane deformation of the layered structure, but also out-of-plane
displacements of the Se atoms and a charge transfer between the two layers.

For a monolayer, the situation is different. The lowest energy supercell is
now $4\times1$ with a 1.19 mRyd/Nb energy gain with respect to the undistorted
cell, followed by $3\times1$. 
We also checked the $2\times1$ and $2\times2$ supercells, which converged to
 the undistorted structure, and the $3\times3$ one, which converged to the 
 same
structure as $3\times1$. This is again consistent with the calculated phonon
spectra. In the ground state Nb atoms are trimerized, qualitatively different
from the bulk CDW. Finally, the bilayer behaves just like a monolayer, with a
$4\times1$ ground state similar to that found for a monolayer.

Having obtained the distorted structure both for the bulk and for
a monolayer, we investigated their electronic properties. We found (Fig.
\ref{fig:dos}) that in the bulk case the density of states (DOS) at the Fermi
level in the distorted $3\times1$ supercell is not much reduced in the CDW,
contrary to what one would expect in the Peierls model, and reconfirming that
the energy gain in this case is not coming from the states in the immediate
vicinity of the Fermi level\cite{Johannes06} but from the occupied Nb 
bands. In the monolayer $4\times1$ supercell the Peierls model is again
violated due to the important role of the electron-phonon matrix element,
however the FS is strongly gapped (see Fig. \ref{fig:monofs} right) 
and displays a clear semimetallic 
behavior in qualitative agreement with
experimental data \cite{Novoselov05}.

\begin{figure}[ptb]
\includegraphics[width=0.9\columnwidth]{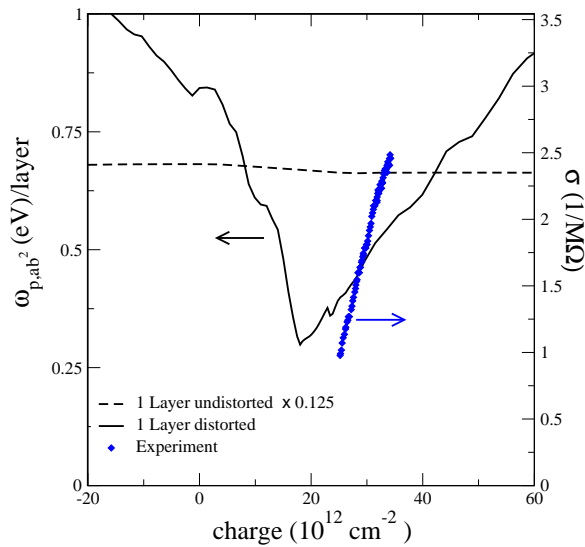}\caption{ (color
online): LAPW plasma frequency and conductivity as a function of 
the gate-induced charge for a single-layer NbSe$_{2}$. The constant relaxation time
approximation is adopted. Experimental data \cite{Novoselov05} are plotted
assuming the charge neutrality at $\approx 30 \times 10^{12}$ cm$^{-2}$. 
The gate voltage for the standard configuration of a 300 nm SiO$_{2}$ 
field-effect-transistor can be obtained from the $x$ axis dividing 
$7.2\times10^{-2}$ cm$^{-2}$ V$^{-1}$. The charge is obtained from the integrated
DOS}%
\label{fig:w_p}%
\end{figure}More insight on the conducting nature of the monolayer is obtained
from the conductivity per layer, namely
\begin{align}
\sigma_{x}^{2D}(E)=\frac{e^{2}}{3 \Omega_{2D} N_{k}}\sum_{\mathbf{k}%
i}v_{\mathbf{k}i,x}^{2}\tau(\mathbf{k}) \delta(\epsilon_{\mathbf{k}i}-E).
\label{eq:sigma}%
\end{align}
where $\Omega_{2D}$ is the area of the two dimensional unit cell of a
NbSe$_{2} $ layer, $e$ is the electron charge, $N_{k}$ the number of k-points
used in the calculation, $\mathbf{v}_{\mathbf{k}i}$ the electron velocity, $E$
the energy. In general $\sigma_{x}^{2D}(E)=\sigma_{x}^{3D}(E) \times L$, where
$L=6.275\mathring{A}$ is the layer thickness. Assuming, $\tau(\mathbf{k}%
)\approx\tau(\epsilon_{f})$, we have that $\sigma_{ab}^{2D}(E)=\tau
(\epsilon_{f})\omega_{p,ab}^{2}/4\pi$.

The results of the $\omega_{p,ab}^{2}$ calculation using the \textit{optics} code
of the WIEN2k package are plotted in Fig. \ref{fig:w_p} for the
monolayer NbSe$_2$ as a function of the gate-induced charge as obtained from the
calculated DOS (Fig. \ref{fig:dos}). For
the undistorted structures, $\omega_{p,ab}$ is essentially constant,
in disagreement with the experiment. On the contrary, in the CDW state
the monolayer plasma frequency shows substantial
variation near $\epsilon_F$, due of its semimetallic
structure. A pseudogap occurs just above $\epsilon_{f}$, so that the
conductivity grows with the gate voltage. In the
constant relaxation-time approximation, and assuming an electron
self-doping of $\approx30\times10^{12}$cm$^{-2}$ we obtain reasonable
agreement with experimental conductivity as 
function of the gate voltage\cite{Novoselov05}(see Fig. \ref{fig:w_p}). 
For a quantitative comparison with experiments, 
however, one would need to know the energy and momentum dependence of the
relaxation time. 
%The large variation of the conductivity as a
%function of gate voltage found in experiments is then a pure manifestation of
%the opening of a pseudogap due to the CDW instability.
%The presence of a self-doping is justified by the following facts. If, as
%it happens in the bulk, the CDW is slightly incommensurate, then our commensurate
%structure will not resolve correctly the position of the pseudogap and a shift of
%the Fermi level is likely. Furthermore, NbSe$_{2}$ is a ionic compound, so
%that one should expect larger self-dopings then in graphene.

To summarize, we have shown that the density functional theory accurately
describes the CDW instability in the bulk 2H-NbSe$_{2},$ and predicts a similar
instability to occur in mono and bilayer. However, while in the bulk the CDW
occurs at $\mathbf{q}_{CDW}\approx(\frac{1}{3},0,0)\frac{2\pi}{a}$, in the
monolayer and bilayer the CDW vector is $\approx(\frac{1}{4},0)\frac{2\pi}{a}%
$. In all these systems the CDW is driven by an enhancement of the
electron-phonon coupling at $\mathbf{q}\sim\mathbf{q}_{CDW}$. Our work solves
the long standing controversy on the origin of CDW in 2H-NbSe$_{2}$ and in
transition metal dichalcogenides.

Unlike the bulk case, when the system remains a good metal in the CDW state,
in a monolayer CDW produces a semimetallic state. This leads
to a large variations of the conductivity 
as a function of the gate voltage, contrary to the undistorted structure and in agreement
with the experiment data\cite{Novoselov05}. Our
calculation shows that a proper description of the CDW instability in reduced
dimension is mandatory to interpret conduction data for low-dimensional
transition-metal-dichalcogenides and opens the way to theoretical
understanding of CDW-based field-effect-devices.

Calculations were performed at the IDRIS supercomputing center (project
081202). 
%MC acknowledges useful discussions with S. Borisenko.


\begin{thebibliography}{99}                                                                                               %


\bibitem {book}G. Gr\"{u}ner, Density Waves in Solids (Addison-Wesley,
Reading, PA, 1994).

\bibitem {Peierls}R. E. Peierls, Quantum theory of solids, Clarendon, Oxford (1955)


\bibitem {Johannes08}M. D. Johannes and I. I. Mazin, Phys. Rev. B \textbf{77},
165135 (2008)

\bibitem {Doran78a}N. J. Doran, B. Ricco, M. Schreiber, D. Titterington and G.
Wexler, J. Phys. C, \textbf{11} 699 (1978)
%\bibitem {Wilson74}J. A. Wilson, F. J. Di Salvo, and S. Mahajan, Phys. Rev.
%Lett. \textbf{32}, 882 (1974), Adv. in Phys., \textbf{24}, 117 (1975)


%\bibitem {Shen08}D. W. Shen \textit{et al.}, Phys. Rev. Lett. \textbf{99}, 216404 (2007)

\bibitem {Johannes06}M. D. Johannes, I. I. Mazin and C. A. Howells, Phyhs.
Rev. B \textbf{73}, 205102 (2006)

\bibitem {Whango92}M. H. Whangho and E. Canadell, J. Am. Chem. Soc.
\textbf{114}, 9587 (1992).

\bibitem {Doran78b}N. J. Doran, J. Phys. C, \textbf{11}, L959 (1978)

\bibitem {Moncton75_77}D. E. Moncton, J. D. Axe and F. J. Salvo Phys. Rev.
Lett. \textbf{34}, 734 (1975) and Phys. Rev. B \textbf{16}, 801 (1977)

\bibitem {Murphy05_08}B. M. Murphy \textit{et al.} Phys. Rev. Lett.
\textbf{95}, 256104 (2005), J. Phys. Cond. Matt. \textbf{20}, 224001 (2008)

\bibitem {Novoselov05}K. S. Novoselov \textit{et al.}, PNAS \textbf{102},
10451 (2005)

%\bibitem {Ayari07}A. Ayary, E. Cobas, O. Ogundadegbe, and M. S. Fuhrer, J.
%App. Phys \textbf{101}, 014507 (2007)

%\bibitem {Pozdorov}V. Pozdorov, M. E. Gershenson, Ch. Kloc, R. Zeis and E.
%Bucher, Appl. Phys. Lett. \textbf{84},3301 (2004)

%\bibitem {Ahn06}C. H. Ahn \textit{et al.}, Rev. mod. Phys. \textbf{78}, 1185 (2006)

%\bibitem {Frindt72}R. F. Frindt, Phys. Rev. Lett. \textbf{28} (1972)

\bibitem {Lebegue09}Leb\`egue and O. Eriksson, Phys. Rev. B \textbf{79},
115409 (2009)

\bibitem {footnote}
Calculations were performed using the QUANTUM-ESPRESSO\cite{QE}
pseudopotential code and the WIEN2K\cite{Wien2K} all-electron LAPW package with the
generalized gradient approximation. For Nb (Se) we use ultrasoft
\cite{Vanderbilt} (norm-conserving \cite{Troullier}) pseudopotentials
including semicore states as valence.

\bibitem {QE}P. Giannozzi \textit{et al.}, J. Phys. Cond. Matt. \textbf{21},
395502 (2009), http://www.quantum-espresso.org

\bibitem{Wien2K} P. Blaha, K. Schwarz, G.K.H. Madsen, D. Kvasnicka and J. Luitz,
WIEN2k, An Augmented Plane Wave +
Local Orbitals Program for Calculating Crystal Properties, Karlheinz
Schwarz, Techn. Universit$\ddot{a}$t Wien, Austria, 2001

%\bibitem {WIEN2K}


%\bibitem {PBE}J.P.Perdew, K.Burke, M.Ernzerhof, Phys. Rev. Lett. \textbf{77},3865 (1996)

%\bibitem {Marezio72}M. Marezio, P. D. Dernier, A. Menth, and G. W. Hull, JR,
%J. Sol. State Chem \textbf{4}, 425 (1972)

\bibitem {Mattheiss73}L. F. Mattheiss, Phys. Rev. Lett. \textbf{30}, 784
(1973), Phys. Rev B \textbf{8}, 3719 (1973)


\bibitem {Inosov08}D. S. Inosov \textit{et al.}, New J. of Phys. \textbf{10},
125027 (2008)
\bibitem {CalandraCaC6}M. Calandra and F. Mauri, Phys. Rev. Lett. \textbf{95},
237002 (2005)
\bibitem {RiceScott75}T. M. Rice and G. K. Scott, Phys. Rev. Lett.,
\textbf{35}, 120 (1975)

\bibitem {Wilson77}J. A. Wilson , Phys. Rev. B, \textbf{15}, 5748 (1997)

\bibitem {Straub99}Th. Straub, $et$ $al$,
% Th. Finteis, R. Claessen, P. Steiner, S.
%H\"ufner, P. Blaha, C. S. Oglesby, and E. Bucher,
 Phys. Rev. Lett.
\textbf{82}, 4504 (1999)


\bibitem {Vanderbilt}D. Vanderbilt, PRB \textbf{41}, 7892 (1990)

\bibitem {Troullier}N. Troullier and J. L. Martins, Phys. Rev. B \textbf{43},
1993 (1991).
%\bibitem {Edwards}J. Edwards and R. F. Frindt, J. Phys. Chem. Solids
\textbf{32}, 2217 (1971)
\end{thebibliography}
\end{document}